\documentclass[12pt]{article}
\usepackage{graphicx}
\usepackage{epstopdf}
\usepackage{wrapfig}

\newcommand{\etal}{{\it et al.}}
\newcommand{\munu}{\mu^-\overline{\nu}}
\newcommand{\Bs}{\overline{B}_s^0}

\newcommand\pubdate{\today}

\def\napoli{Department of Physics\\
Syracuse University, Syracuse, NY, United States of America}
\def\Title#1{\begin{center} {\Large #1 } \end{center}}
\def\Author#1{\begin{center}{ \sc #1} \end{center}}
\def\Address#1{\begin{center}{ \it #1} \end{center}}

\newcommand\pubblock{\rightline{\begin{tabular}{l}% \pubnumber\\
         \pubdate  \end{tabular}}}
\newenvironment{Abstract}{\begin{quotation}  }{\end{quotation}}
\newenvironment{Presented}{\begin{quotation} \begin{center} 
             PRESENTED AT\end{center}\bigskip 
      \begin{center}\begin{large}}{\end{large}\end{center} \end{quotation}}

\def\beq{\begin{equation}}
\def\eeq#1{\label{#1}\end{equation}}
\def\eeqn{\end{equation}}

\def\beqa{\begin{eqnarray}}
\def\eeqa#1{\label{#1}\end{eqnarray}}
\def\eeqan{\end{eqnarray}}

\let\bar=\overbar

\def\etal{{\it et al.}}

\def\Dslash{\not{\hbox{\kern-4pt $D$}}}
\def\dslash{\not{\hbox{\kern-2pt $\del$}}}

\def\msb{{\bar{\ssstyle M \kern -1pt S}}}

\begin{document}
\begin{titlepage}
\pubblock

\vfill
\Title{Semileptonic $B_s$ decay measurements at LHCb}
\vfill
\Author{ Phillip Urquijo}
\Address{\napoli}
\vfill
\begin{Abstract}
I discuss the first semileptonic $B_s$ decay measurements performed at LHCb, and the prospects for precise studies of $B_s$ semileptonic decay properties.
The analyses covered here are based on data collected with the LHCb detector in proton-proton collisions at a centre-of-mass energy of 7 TeV. In particular the semileptonic decays  $\Bs\to D_s^+ X \munu$ and $\Bs\to D^0K^+ X \munu$ are studied. Two structures are observed in the $D^0K^+$ mass spectrum at masses consistent with the known $D_{s1}(2536)^+$ and $D^*_{s2}(2573)^{+}$ mesons. The measured branching fractions relative to the total $\Bs$ semileptonic rate are
${{\cal{B}}(\Bs\to D_{s2}^{*+} X \munu)}/{{\cal{B}}(\Bs\to X \munu)}= (3.3\pm 1.0\pm 0.4 )\%$, and
${{\cal{B}}(\Bs\to D_{s1}^+ X \munu)}/{{\cal{B}}(\Bs\to X \munu)}= (5.4\pm 1.2({\rm stat})\pm 0.5({\rm sys}))\%$.
This is the first observation of the $D_{s2}^{*+}$ state in $\Bs$ decays.
\end{Abstract}
\vfill
\begin{Presented}
CKM2010\\
6th International Workshop on the CKM Unitarity Triangle University of Warwick, UK, 6-10 September 2010
\end{Presented}
\vfill
\end{titlepage}
\def\thefootnote{\fnsymbol{footnote}}
\setcounter{footnote}{0}

\section{Introduction}
Semileptonic decays of $B$ mesons constitute a very important class of decays for determination of the elements of the Cabibbo-Kobayashi-Maskawa (CKM) matrix and  for understanding the origin of CP violation in the Standard Model (SM).   The light $B^0$ and $B^+$ meson decays have been precisely measured by experiment and well studied in theories. The experimental information on the production and decay of the $B_s$ and the $\Lambda_b$ hadrons, however, is relatively limited.  The interest in the physics of the $B_s$ has intensified in recent years, particularly concerning studies of the dilepton production asymmetry in $b\bar b$ production.  At the LHC, besides $B^0$ and $B^+$ mesons, an abundant number of  $B_s$ mesons and $\Lambda_b$ baryons are produced.  The LHCb experiment has already begun collecting large $b$ data  sets, allowing new measurements of semileptonic (SL) decays of $B_s$ and $\Lambda_b$ hadrons~\cite{LHCb-det}, with high sensitivity to low $p_T$.
LHCb first exploited SL $b$ decays to determine the inclusive $b\bar b$ production cross section \cite{1stpaper}. The decay mode of $b \to D^0 \mu X$ was reconstructed to determine the number of produced $b$'s.  
In this proceeding we focus on studies of $\Bs$ SL decays, recently performed by LHCb~\cite{ds2paper}.  In SL $\Bs$ decays, the Cabibbo favoured $b\to c$ transition results in a single charm hadron, which can be a $D_s^+$, a $D_s^{*+}$ or another excited $c\overline{s}$ state. The excited  $c\overline{s}$ meson modes may disintegrate into final states containing either $D_s$ or $DK$.  One such $c \bar s$ state is the $D_{s1}^+$, that decays into $D^*K$, and another is the $D_{s2}^{*+}$, that has been observed to decay directly into $DK$ \cite{PDG}.  The relative proportion of these final states provides essential information on the structure of $B_s$ SL decays. 
The analysis discussed here uses a data sample of approximately 20 pb$^{-1}$  collected from 7 TeV centre-of-mass energy $pp$ collisions at the LHC during 2010. For the first 3 pb$^{-1}$ of these data a single muon trigger was used, which was not available for the remainder of data taking. This sample is well suited to determine the number of SL $\Bs$ decays, that we take as the sum of $D_s^+ X\munu$, $D^0K^+X\munu$ and $D^+K^0 X\munu$ decays, ignoring charmless $\Bs$ decays  ($\approx$1\%). The entire 20  pb$^{-1}$  sample is useful
for establishing signal significance, resonance parameter determination, and the ratio of numbers of events in the $D^0K^+$ states.Later in the proceeding we discuss future prospects for precision measurements of HQET and CKM parameters, $|V_{ub}|/|V_{cb}|$.

\section{Semileptonic $B_s$ decay reconstruction}
We first select a charm hadron that forms a vertex with a muon.  We consider two cases: (i)  $D_s^+\to K^+K^-\pi^+$, used to normalize the $\Bs$ yield; and (ii) $D^0\to K^-\pi^+$ decays combined with an additional $K^+$  that forms a  vertex with the $D^0$ and the $\mu^-$ in order to search for $\Bs$ SL decays that might occur via an excited  $c\overline{s}$ meson that decays into $D^0K^+$. 
Most charm hadrons  are produced directly via $pp\to c\overline{c} X$ interactions at the LHC and are denoted as ``Prompt".  Charm is also produced in $pp\to b\overline{b} X$ collisions where the $b$ hadron decays into charm,  called charm from $b$ hadrons or ``Dfb". Muon candidates used in the first 3 pb$^{-1}$ sample must have triggered the event and have $p_{\rm T}>$ 1200 MeV.
The  criteria for $D_s^+$ and $D^0$ mesons include identifying K and $\pi$ candidates using the RICH.  We require that the $p_{\rm T}$ of the $K$'s and $\pi$ be $>$300 MeV, and that their scalar sum be $>$2100 MeV ($D_s^+)$ or $>$1400 MeV ($D^0$).
Since charm mesons travel before decaying, the $K$ and $\pi$ tracks are required to not point to the  primary vertex. The $K$ and $\pi$ candidate tracks must also be consistent with the charm decay vertex.
The charm candidate's decay vertex must be detached from the interaction point.   $\Bs$ candidates formed from $D^+_s$ and muon combinations must form a vertex, point at the primary vertex, and have an invariant mass in the range 3.10~GeV$<m(D_s^+\mu^-)<5.10$ GeV.
The analysis for the $D_s^+ X\munu$ uses the 3 pb$^{-1}$ sample.
\begin{figure}[bt]\centering
\includegraphics[width=0.8\linewidth]{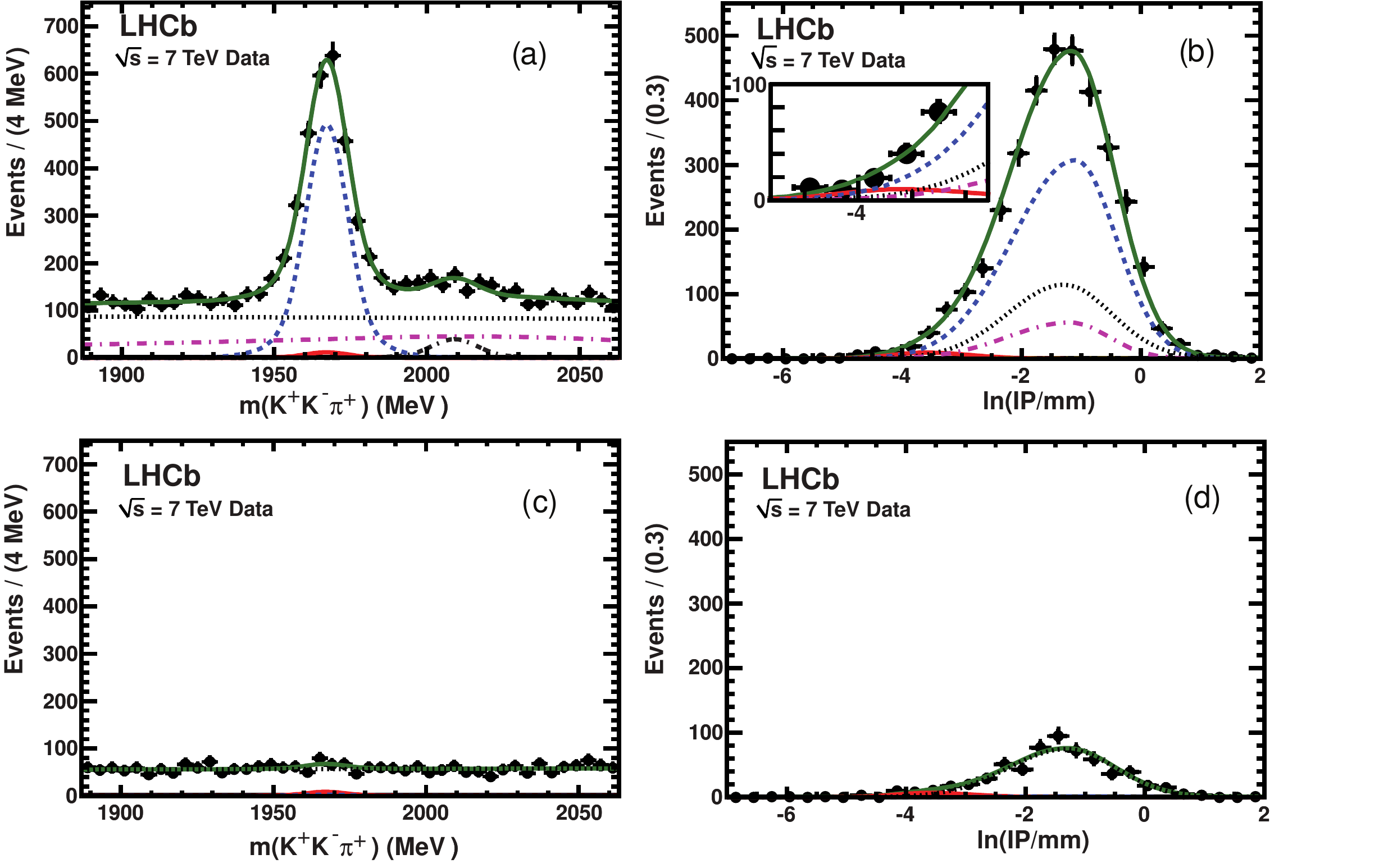}
\caption{The invariant $K^+K^-\pi^+$ mass spectra for events associated with a $\mu$ for the 3 pb$^{-1}$ sample with $2<\eta<6$ for RS (a) and WS (c). Also shown is ln(IP) of the $D_s^+$ candidates for (b) RS and (d) WS samples. The curves correspond to total (solid green), Dfb (blue-dashed), sideband (black dotted), misinterpreted $\Lambda_c^+$ (purple-dot-dashed ), $D^{*+}$ (black dash-dash-dot), and prompt (solid red). 
 } \label{Ds-overall-Lc}
\end{figure}
The $K^+K^-\pi^+$ mass spectra for both the right-sign  (RS $K^+K^-\pi^+$ + $\mu^-$) and wrong-sign (WS $K^+K^-\pi^+$ + $\mu^+$) candidates, as well as the ln(IP/mm)($D_s^+$) distributions for events with mass combinations within $\pm$20 MeV of the $D_s^+$ mass are shown in Fig.~\ref{Ds-overall-Lc} for $2<\eta<6$.  We perform unbinned extended maximum likelihood fits to the 2-D distributions in $K^+K^-\pi^+$ invariant mass and ln(IP/mm).  
The prompt IP shape is determined from directly produced charm \cite{1stpaper}, and MC is used for the Dfb shape.  The fit separates
contributions from Dfb, prompt, and false combinations. Background components for $D^{*+}\to \pi^+D^0\to K^+K^-\pi^+$ and the reflection from  $\Lambda_c^+\to pK^-\pi^+$ decay,  are also included. 
The $D_s^+$ fits and efficiency corrections are performed in $\eta$ bins to avoid $\eta$ dependent data/MC differences.
The procedure yields  2233$\pm$60 RS Dfb events in the $D_s^+ X \munu$ channel for $2<\eta(b)<6$, uncorrected for efficiency. This yield is further reduced by 5.1\% for $b$ backgrounds.

\section{\boldmath Measurement of $D^0 K^+ X\munu$}
\begin{wrapfigure}{r}{0.48\textwidth}\centering
\centering
\includegraphics[width=0.48\textwidth]{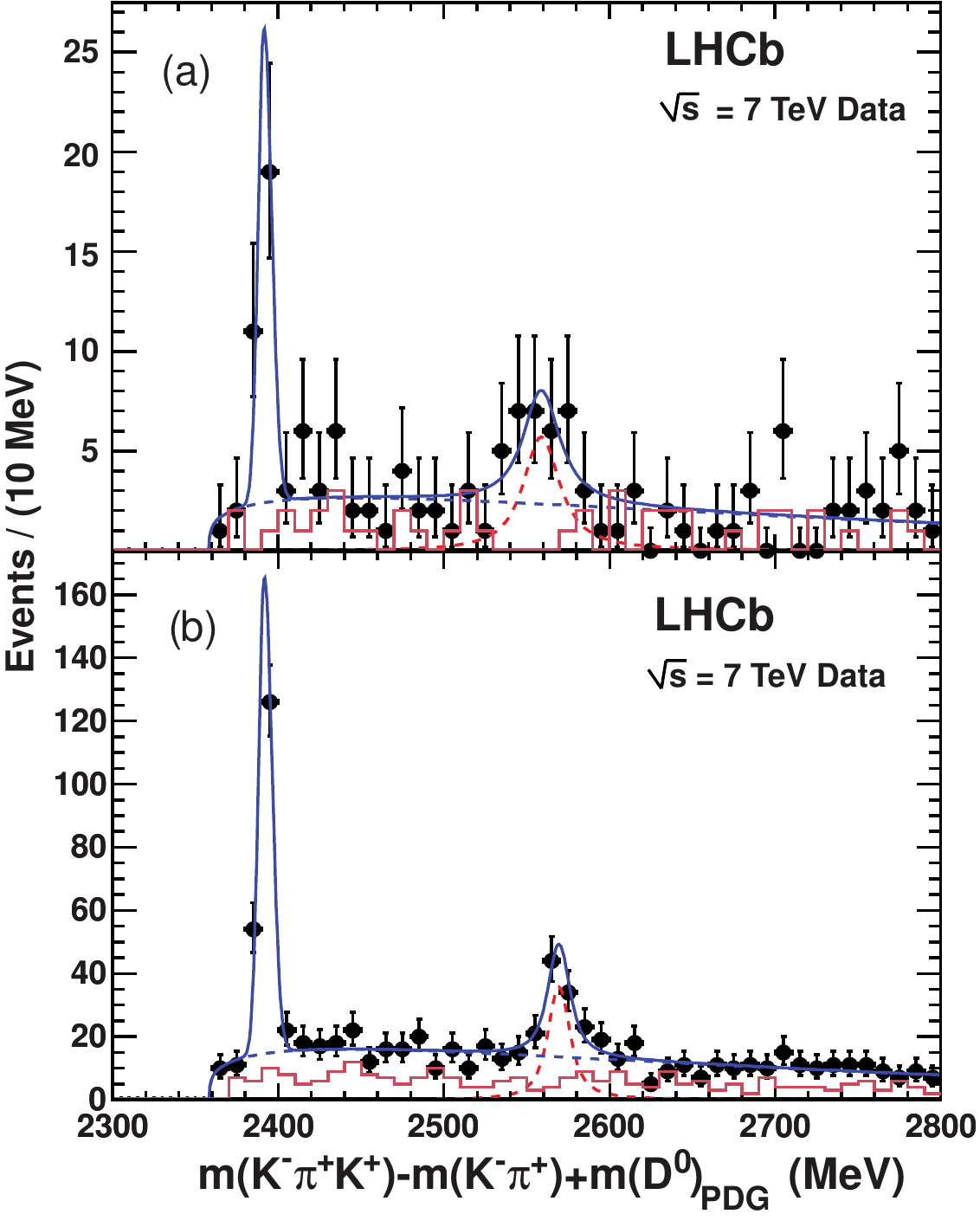}
\caption{$m(K^-\pi^+K^+)-m(K^-\pi^+)+m(D^0)_{\rm PDG}$ for events with $m(K^-\pi^+)$  within $\pm$20 MeV of the $D^0$ mass (black points) in SL decays, with the histogram showing WS $K$ events. (a) 3 pb$^{-1}$ data sample, (b) 20 pb$^{-1}$ sample.
 } \label{m_d0k_vetop}
\end{wrapfigure}
Semileptonic decays of $\Bs$ mesons usually result in a $D_s^+$ meson in the final state. The SL decay may also proceed via a $c\overline{s}$ excitation,  which can decay to either $D K$ or $D^* K$ resonances, or non-resonant $DK$.  To search for these final states, we measure the $D^0 K^+ X \munu$ yield. 
To seek events with a $D^0$ candidate and an additional $K^+$ we require that
the $K^+$ candidate has $p_{\rm T} > 300$ MeV, be identified in the RICH, not point to the IP, and that the vector sum $p_{\rm T}$  of the $D^0$ and $K$ be $>1500$ MeV. The $B$ candidate must have a mass in the range 3.09~GeV$<m(D^0K^+\mu^-)<5.09$ GeV, form a vertex and point at the primary vertex. 
Figure~\ref{m_d0k_vetop}(a) shows the $D^0 K^+$ invariant mass spectrum in the 3 pb$^{-1}$ sample. 
A signal near threshold is seen corresponding to the $D_{s1}(2536)^+$, but at a lower mass of 2392 MeV as it decays via $D^{*0}K^+$, and we do not reconstruct the $\gamma$ or $\pi^0$ from the $D^{*0}$.  
This final state was seen previously with $D_{s1}^{+}\to D^{*+}K_S^0$ decays \cite{D0}. There is also a feature near the mass of the $D^*_{s2}(2573)^{+}$ meson. 
To determine the signal yield we perform an unbinned maximum likelihood fit.  We assume that the $\Bs\to D^0K^+X\munu$ signal  is saturated by the $D_{s1}^{+}$ and $D_{s2}^{*+}$ states. For the $D_{s1}^+$ we find  24.4$\pm$5.5 $D_{s1}^+$ events. 
To confirm the $D_{s2}^{*+}$ signal we use the full data sample of 20 pb$^{-1}$, in which we accept all triggered events. This sample is useful to increase statistics, however using multiple triggers makes it more difficult to measure the total number of $\Bs$ decays. The measurement of the relative yields of $D_{s2}^{*+}$ to $D_{s1}^{+}$ is not affected.  Figure~\ref{m_d0k_vetop}(b) shows the $D^0K^+$ invariant mass spectrum. The difference between RS and WS events outside of the resonant peaks is consistent with background from other $b$ decays .
The fit to the $D_{s1}^+$ yields 155$\pm$15 signal events, a $D^0K^+$ mass of 2391.6$\pm$0.5 MeV, and 4.0$\pm$0.4 MeV for the width.
For the $D_{s2}^{*+}$ we find a mass  of 2569.4$\pm$1.6$\pm$0.5 MeV,  a width of 12.1$\pm$4.5$\pm$1.6 MeV, and 82$\pm$17 events. The $D_{s2}^{*+}$ signal significance is 8$\sigma$.
The relative branching fractions (${\cal{B}}$) are determined from the 20 pb$^{-1}$ sample, assuming the $D_{s1}^{+}$ decays only into $D^*K$ final states, the $D_{s2}^{*+}$ decays only into $DK$ final states, and isospin is conserved.  The only observed decays of $D_{s2}^{*+}$ are to $DK$ final states. The $D_{s2}^{*+}/D_{s1}^+$ event ratio is computed as
\begin{equation}
\label{eq:Ds2BR}
{\cal{B}}(\Bs\to D_{s2}^{*+} X \munu)/{\cal{B}}(\Bs\to D_{s1}^+ X \munu)=0.61\pm 0.14\pm 0.05.
\end{equation}
The relative ${\cal{B}}$ of the $D_{s1}^+$ with respect to the total $B_s$ SL rate
is measured using the 3 pb$^{-1}$ sample.
The number of $\Bs$ SL decay events  in this sample is evaluated from the efficiency corrected sum of the $\Bs\to D_s^+ X \munu$  events and twice the efficiency corrected $\Bs\to D^0 X K^+\munu$ yield (accounting for the unmeasured $D^+K^0 X\munu$).
A small component of $B\to D_s^+ K X\munu$ is subtracted~\cite{BtoDsK}, reducing the $D_s^+ X\munu$ yield by 3.2\%. The overall uncertainty on the $\Bs$ SL yield is 6.6\%, mainly from the absolute $D_s^+$ branching ratio (4.9\%), and the uncertainty from the amount of $D^0K^+X\munu$ events (3.0\%). 
The $D_{s2}^{*+}$ ${\cal{B}}$ is computed using this sample and the result from Eq.~\ref{eq:Ds2BR}. The relative ${\cal{B}}$s are:
\begin{eqnarray}
{\cal{B}}(\Bs\to D_{s2}^{*+} X \munu){\cal{B}}/(\Bs\to X \munu)&=& (3.3\pm 1.0\pm 0.4 )\%\nonumber\\
{\cal{B}}(\Bs\to D_{s1}^+ X \munu){\cal{B}}/(\Bs\to X \munu)     &=& (5.4\pm 1.2\pm 0.5)\%,
\end{eqnarray}
where the systematic uncertainties are mainly from the detection efficiency (5\%),  and the number of $\Bs$ SL decays (6.6\%). 
There is an additional systematic uncertainty of 8\% on the  $D_{s2}^{*+}$  yield from the fit. The ${\cal{B}}$ for the relative rate of $D_{s1}^+$ decay is consistent with but smaller than D0\cite{D0}.
These values were predicted in the ISGW2 model as 3.2\% and 5.7\%, for $D_{s2}^{*+}$ and $D_{s1}^{+}$, respectively,  in good agreement with our observations \cite{ISGW2}.  
\section{Prospects}
The measurement of the exclusive decay modes of $B_s\rightarrow D_s \ell^+{\nu_\ell}$ and $B_s\rightarrow D_s^*\ell^+{\nu_\ell}$ have not previously been detected separately.   Information can be determined on the contribution from $D_s$ and $D^*_s$ and excited $c \bar s$ states to the $B \to D_s \mu \nu X$ final state by reconstructing the rest frame observables. In LHCb these can be accessed by determining the $B$ flight direction vector from the separation of the primary (IP) and secondary (DfB + $\mu$) vertices, and solving for the neutrino momentum with a two-fold ambiguity. The kinematic resolution of e.g. $q^2$ is similar to \begin{wrapfigure}{r}{0.44\textwidth}\centering
\centering
\includegraphics[width=0.44 \textwidth]{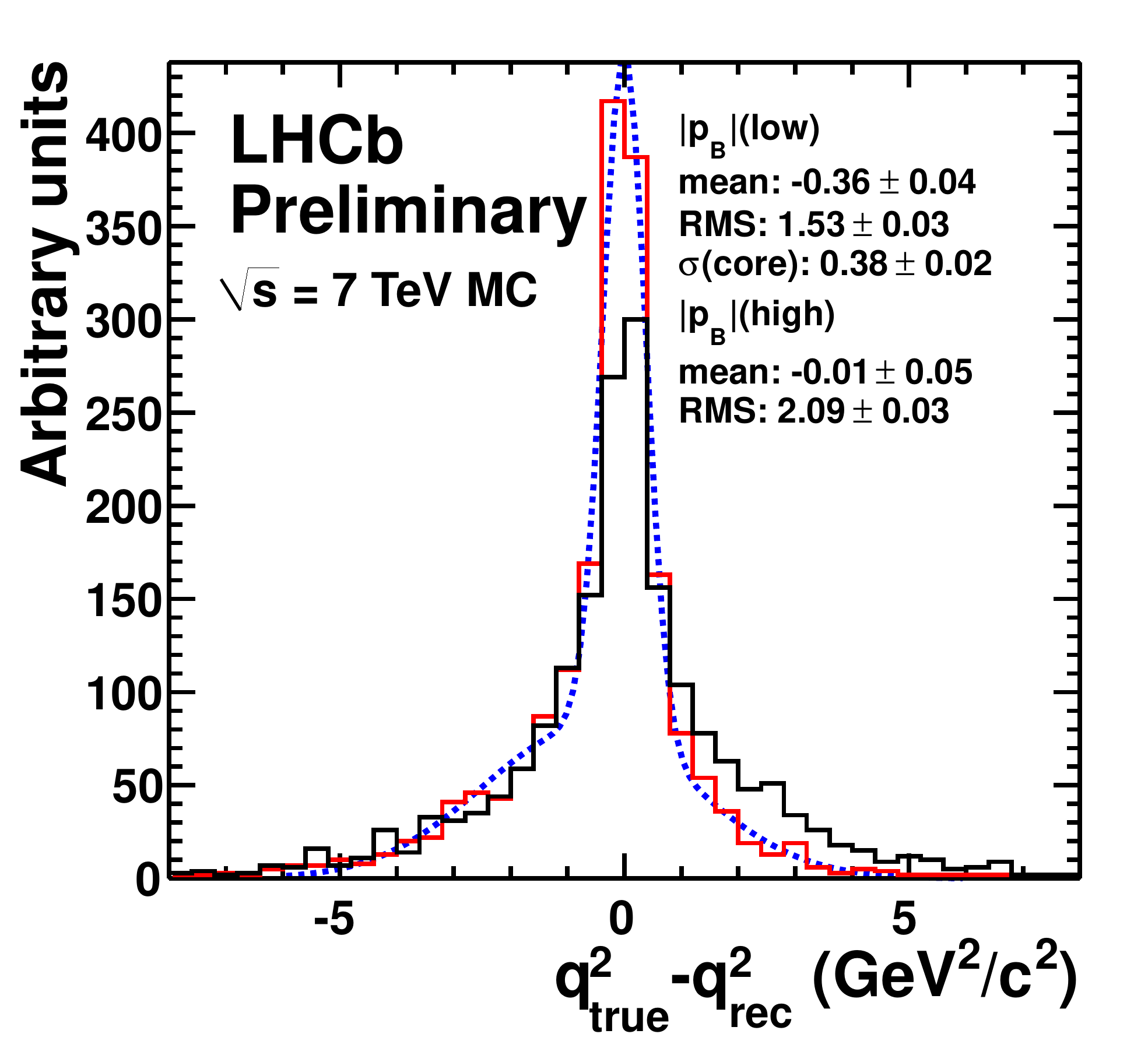}
\caption{$q^2$ resolution with the neutrino reconstruction technique.} \label{nures}
\end{wrapfigure}
that observed in the $B-$factories, and with sufficient phase space for reliable comparison to theory. The corresponding resolution in the $D_s$ mode is shown in Fig. \ref{nures}.  Preliminary results indicate promise for measuring ${\cal{B}}$s, and form factor information in both $B_s$ and $\Lambda_b$ decays, with the potential to study Cabibbo suppressed modes.
The ultimate goals will be measurements of the CKM matrix element $|V_{ub}|$ in exclusive $B_s$ decays, and the dilepton asymmetry in $B_s$ decays.
Input from lattice is still required for high precision interpretation of $B_s$ decays,  for both, Cabibbo favoured and suppressed modes.

\section{Conclusions}
LHCb has exploited semileptonic $B$ decays to measure the $b$ cross section at the LHC, and perform a preliminary measurement of the $b$-quark hadronisation fractions. LHCb has demonstrated unprecedented sensitivity to exclusive $B_s$ SL modes, and plans for improved precision studies of the full SL decay width using neutrino reconstruction, as well as analogous studies in the $\Lambda_b$ modes.  The first observation has been made of  the SL decay $\Bs\to D^*_{s2}(2573)^{+}  X \munu$ and its ${\cal{B}}$ relative to the total SL $\Bs$ decay rate has been measured.
The same ratio with $\Bs\to D_{s1}(2536)^+ X \munu$ SL decays ratio has also been measured improving upon previous levels of precision.

\end{document}